\centerline{\bf Anti de Sitter Gravity from BF-Chern-Simons-Higgs
Theories}
\bigskip
\centerline {Carlos Castro}
\centerline {Center for Theoretical Studies of Physical Systems}
\centerline{Clark Atlanta University, Atlanta, GA. 30314}
\bigskip
\centerline{\bf Abstract}
\bigskip
It is shown that an action inspired from a BF and Chern-Simons model,
based on the $AdS_4$ isometry group SO(3, 2), with the inclusion of a
Higgs potential term, furnishes the MacDowell-Mansouri-Chamseddine-West
action for gravity, with a Gauss-Bonnet and cosmological constant term.
The $AdS_4 $ space is a natural vacuum of the theory. Using Vasiliev's
procedure to construct higher spin massless fields in AdS spaces and a
suitable star product, we discuss the preliminary steps to construct the
corresponding higher-spin action in $AdS_4$ space representing the higher
spin extension of this model.  Brief remarks on Noncommutative Gravity
are made.
\bigskip
It has been known for some time that the first order
formalism of pure $3D$ gravity is a $BF$ theory. The same
occurs for $4D$ gravity and higher-dim gravity if
additional quadratic constraints on the $B$ field are added
to the Lagrangian [1]. These $BF$ theories are very
suitable for the spin-foam quantization techniques [2]
that are valid in any dimension due to common structures to
all constrained $BF$ theories. A very important result is
that $4D$ YM theory can be obtained from a deformation of a
Topological $BF$ theory [3].
Deformations of a world volume $BF$ theory as possible
deformations of a Topological open membrane model by means
of the antifield $BRST$ formalism was performed by Ikeda
[4]. Noncommutative structures on the boundaries of the
open membrane were obtained as a generalization of the path
integral representation of the star product deformation
given by Kontsevich [5]. The path integral representation
of the Kontsevich formula on a Poisson manifold was given
as a perturbative expansion of a two-dim field theory
defined on the open two-dim disk $D^2$ [6].
Similar star product structures appear in open string
theory with a nonzero Neveu-Schwarz $B$ field [7]. Based on
the BF-YM relation [3], and our branes-YM relation [8]
based on a Moyal deformation quantization of Generalized
Yang-Mills, we are going to close the triangle
BF/YM/Gravity by showing in this work how gravity with
a negative cosmological constant can be obtained from a
BF-CS-Higgs theory. The $4D$ action below is inspired from
a BF-CS model defined on the boundary of the $5D$ region $D^2 \times R^3
$,
where $D^2$ is the open domain of the
two-dim disk. $AdS_4$ has the topology of $ S^1(time)\times R^3 $ which
can
be seen as the (lateral) boundary of $ D^2 \times R^3 $.
The relevant BF-CS-Higgs $inspired$ action is based on the
isometry group of $AdS_4$ space given by $SO(3, 2)$, that
also coincides with the conformal group of the $3$-dim
boundary of $ AdS_4$: $ S^1 \times S^2 $. The action
involves the gauge fields $A^{AB}_\mu $ and a family of
Higgs scalars $\phi^A~$ that are $SO(3,2)$ vector-valued
$0$-forms and the indices run from $ A = 1,2,3,4,5$.
The action can be written in a compact notation using
differential forms:
$$S_{BF-CS-Higgs} = \int_{M_{4}} \phi \wedge F \wedge F + \phi\wedge d_A
\phi \wedge d_A \phi \wedge d_A \phi \wedge d_A \phi - V_H (\phi).
\eqno (1) $$
A word of caution: strictly speaking, because we are using
a covariantized exterior differential $d_A$, we don't have
the standard BF-CS theory. For this reason we use the
terminology BF-CS-Higgs inspired model.
The $5D$ orgins of the BF-CS inspired action is of the form
$$\int_{D^2 \times R^3} d\phi \wedge F \wedge F \leftrightarrow \int B
\wedge F_{4}. ~~~ B = d\phi.~~~F_{4} = F \wedge F. \eqno (2) $$
$$ \int _{D^2 \times R^3} d \phi\wedge d \phi \wedge d \phi \wedge d
\phi \wedge d\phi \rightarrow \int_{S^1 \times R^3} \phi \wedge d \phi
\wedge d \phi \wedge d \phi \wedge d\phi. \eqno (3) $$
The $F$ and $ F_{4} = F \wedge F$ fields satisfy the Bianchi-identity:
$$ F = d_A A = d A + A \wedge A.~~~d^2_A \phi = F \phi \not=0. ~~~ d^2_A
A = d_A F = 0 \Rightarrow d_A (F \wedge F) = 0. \eqno (4) $$
The Higgs potential is:
$$ V_H(\phi) = \kappa_1 (\eta_{AB} \phi^A \phi^B - v^2)^2. ~~~
\eta_{AB} = (+, +, +, -, -). ~~~\kappa_1 = constant. \eqno (5) $$
The gauge covariant exterior differential is defined: $ d_A = d + A $
so that $d_A \phi = d\phi + A \wedge \phi $ and the $SO(3, 2)$ field
strengths:
$ F = dA + A \wedge A $
are the usual ones associated with the $SO(3, 2) $ gauge fields in the
adjoint representation:
$$ A^{AB}_\mu = A^{ab}_\mu~;~ A^{5a}_\mu; ~ a, b = 1,2,3,4. \eqno (6)$$
which, after symmetry breaking, will be later identified as the Lorentz
spin connection $ \omega^{ab}_\mu $ and the vielbein field respectively:
$ A^{5a}_\mu = \lambda e^a_\mu $ where $ \lambda $ is the inverse
$AdS_4$ scale.
The Lie algebra $SO(3, 2) $ generators obey the commutation relations:
$$ [ M_{AB}, M_{CD} ] = \eta_{BC} M_{AD} - \eta_{AC} M_{BD} + \eta_{AD}
M_{BC} - \eta_{BD} M_{AC}. \eqno (7) $$
We will show next how gravitational actions with a
cosmological constant can be obtained from an action
$inspired$ from a BF-CS-Higgs theory. Before we begin with
our derivation we must emphasize that our procedure,
although very similar in many respects to Wilczek's work
[9], $differs$ from his approach in several aspects.
1- Our action given by eq-(4) is not the $same$ as
Wilczek's action. We have a covariantized Chern-Simons term
instead of a Jacobian-squared expression and it is not
$necessary$ to choose a preferred volume [1], leaving a
residual invariance under volume-preserving
diffeomorphisms, in order to retrieve the
MacDowell-Mansouri-Chamseddine-West (MMCW) action for
gravity [14].
2-Our procedure is tightly connected to the the topological
$BF$ origins of ordinary gravity [2] and of Yang-Mills
theories [3,4]. The connection to $ BF$ and Chern-Simons
theories was overlooked in [9].
3- A variation of our action (a minimization of the Higgs
potential) with respect to the scalar fields $ \phi^A$
allows to eliminate them from the action and to generate
the MMCW action for gravity [14], after an spontaneous symmetry
breaking of the Anti de Sitter group $SO(3, 2)$ symmetry
down to the Lorentz $ SO(3,1)$. The latter MMCW action
admits $ AdS_4$ as the natural vacuum solution $F^{AB}_{\mu\nu}= 0 $ (the
MMCW action for $ AdS_4$ space is naturally zero).
Whereas in the approach of [9], a $simultaneous$
minimization procedure of the Higgs potential $and$ the
Jacobian-squared term (by choosing a preferred volume) will
automatically $constrain$ the action to $zero$ because a
variation of the Wilczek action w.r.t the scalars will then
constrain the $ \phi \wedge F \wedge F $ terms to zero if
both the Higgs potential and the Jacobian-squared terms are
stationarized. This does not occur in our case since we
have a different action than [9]. In our case, the action
is zero for the $ AdS_4$ vacuum solution of the MMCW model.
The Higgs potential is minimized at tree level when the
$vev$ are:
$$ < \phi^ 5 > = v. ~~~ < \phi^a > = 0. ~~~ a = 1, 2, 3, 4.. \eqno (8)
$$ which means that one is freezing-in at each spacetime point the
internal $ 5$ direction of the internal space of the group
$ SO(3, 2)$. Using these conditions (8) in the definitions
of the gauge covariant derivatives acting on the internal
$SO(3,2)$-vector-valued spacetime scalars $\phi^A (x) $, we
have that at tree level:
$$ \nabla_\mu \phi^5 = \partial_\mu \phi^5 + A^{5a}_\mu \phi^a = 0. ~~~
\nabla_\mu \phi^a = \partial_\mu \phi^a + A^{ab}_\mu \phi^b + A^{a5}_\mu
\phi^5 = A^{a5}_\mu v. \eqno (9) $$
A variation of the action w.r.t the scalars $\phi^a$ yields the zero
Torsion condition after imposing the results (8, 9) solely $after$ the
variations have been taken place.
Therefore it is not necessary to impose by hand the zero torsion
condition like in the MMCW procedure. Varying w.r.t the $\phi^a$ yields
the $ SO(3, 2)$-covariantized Euler-Lagrange equations that lead
naturally to the zero Torsion $ T^a_{\mu\nu} $ condition:
$$ {\delta S \over \delta \phi^a} - d_A {\delta S \over \delta (d_A
\phi^a)} = 0 \Rightarrow F^{5a}_{\mu\nu} = T^a_{\mu\nu} = \partial_\mu
e^a_\nu + \omega^{ab}_\mu e^b_\nu - \mu \leftrightarrow \nu = 0
\Rightarrow \omega^{ab}_\mu = \omega (e^a_\mu). \eqno (10) $$
and one recovers the standard Levi-Civita (spin) connection in terms of
the (vielbein) metric. A variation w.r.t the remaining $ \phi^5 $
scalar yields after using the relation $ A^{a5}_\mu = \lambda e^a_\mu $:
$$ F^{ab}_{\mu\nu} F^{cd}_{\rho\tau} \epsilon_{abcd5}
\epsilon^{\mu\nu\rho\tau} + 5 \lambda^4 v^4 e^a_\mu e^b_\nu e^c_\rho
e^d_\tau \epsilon_{abcd5} \epsilon^{\mu\nu\rho\tau} = 0 \leftrightarrow
$$
$$ - {1\over 5} \phi^5 F^{ab}_{\mu\nu} F^{cd}_{\rho\tau}
\epsilon_{abcd5} \epsilon^{\mu\nu\rho\tau} = \phi^5 \nabla_\mu \phi^a
\nabla_\nu \phi^b \nabla_\rho \phi^c \nabla_\tau \phi^d \epsilon_{abcd5}
\epsilon^{\mu\nu\rho\tau} \eqno (11) $$
Using these last equations (8-11), after the minimization procedure,
will allows us to eliminate on-shell all the scalars $\phi^A $ from the
action (4) furnishing the MacDowell-Mansouri-Chamseddine-West action
for gravity as a result of an spontaneous symmetry breaking of the
internal
$SO(3, 2)$ gauge symmetry due to the Higgs mechanism leaving unbroken the
$SO(3, 1)$ Lorentz symmetry:
$$ S_{MMCW} = {4 \over 5} v \int d^4x ~ F^{ab}_{\mu\nu}
F^{cd}_{\rho\tau} \epsilon_{abcd5} \epsilon^{\mu\nu\rho\tau}. \eqno
(12) $$
with the main advantage that it is no longer necessary to
impose by hand the zero Torsion condition in order to
arrive at the Einstein-Hilbert action. On the contrary, the
zero Torsion condition is a direct result of the
spontaneous symmetry breaking and the dynamics of the
orginal BF-CS inspired action. In general, performing the
decomposition
$$ A^{ab} _\mu = \omega^{ab}_\mu. ~~~ A^{a5} _\mu = \lambda e^a_\mu.
\eqno (13) $$.
where $\lambda$ is the inverse length scale of the model (like the
$AdS_4$ scale) and inserting these relations into the MMCW action yields
finally the Einstein-Hilbert action, the cosmological constant plus the
Gauss-Bonnet Topological invariant in $ D = 4 $, respectively:
$$ S = {8\over 5} \lambda^2 v \int R \wedge e \wedge e ~+ {4 \over 5}
\lambda^4 v \int e\wedge e \wedge e \wedge e ~ +{4 \over 5} v \int R
\wedge R. \eqno (14) $$
which implies that the gravitational $L^2_{Planck} $ and the
cosmological constant $\Lambda_c$ are fixed in terms of $\lambda, v $,
up to numerical factors, as:
$$ \lambda^2 v = {1\over L^2_P}. ~~~ \Lambda_c = \lambda^4 v. \eqno (15a)
$$
Eliminating the vacuum expectation value (vev) value $v$ from eq-(15a)
yields a geometric mean relationship among the three scales:
$$ \lambda^2 {1 \over L^2_P} = \Lambda_c \Rightarrow L^4_P \leq {
1\over \Lambda_c} \leq {1 \over \lambda^4}~\Rightarrow ~ {1\over L^4_P}
\geq
{\Lambda_c} \geq {\lambda^4}. \eqno (15b) $$
Hence we have an upper/lower bound on the cosmological constant
$ \Lambda_c $ based on the Planck scale and the
$AdS$ inverse scale $ \lambda $.
We will use precisely these geometric mean relations (15) to get an
estimate of the cosmological constant based on our results [8] on the
relation among deformation quantization, the large $ N$ limit of (
Generalized) Yang-Mills and $p$-branes. $ SU(N)$ reduced and quenched
Yang-Mills have recently been shown by us, via a Moyal deformation
quantization procedure [8], to be related to Hadronic Bags and
Chern-Simons Membranes (dynamical boundaries)in the large $ N $ limit.
In particular, the value of the dynamically-generated bag tension $ T $
was shown to be related to the lattice spacing, $ a $, associated with
the large $ N$ quenched, reduced $ SU(N)$ YM theory as follows:
$$ T = \mu^4 \sim {1 \over a^4 g^2_{YM}}, \eqno (16) $$
where $ g_{YM} $ is the YM coupling constant and $ \mu$ is
the bag constant (mass dimensions). Such results [8] are
compatible with the the Maldacena $ AdS/CFT$ duality
conjecture. Based on the result that a stack of $ N$
coincident $ D3$ branes (whose world-volumes are
four-dimensional), in the large $N$ limit, are related to
black $p = 3$ -brane solutions to closed type $ II~ B $ string
theory in $ D = 10$, and whose near-horizon geometry is
given by $ AdS_5 \times S^5$, one can set the lattice
spacing $a$ associated with the large $N$ quenched, reduced
$SU(N)$ YM in terms of the Planck scale $L_P$ to be:
$$ a^4 = N L^4_P. \eqno (18) $$ which merely sates that the hadronic
bag scale $ a = N^{1/4} L_P$
Inserting this relationship (18) into (17) yields:
$$ T = \mu^4 \sim {1 \over NL^4_P g^2_{YM}} \Rightarrow \mu^{-4} \sim (
N g^2_{YM}) L^4_P, \eqno (19) $$
which has a similar form as the celebrated Maldacena result
relating the size of the $ AdS_5$ throat $\rho^4 $ to the
't Hooft coupling $ N g^2_{YM} $ and the Planck scale $
L^4_P \sim (\alpha')^2 $, the inverse string tension
squared. We believe that this is more than just a mere
numerical coincidence.
Therefore, if one sets the inverse $AdS_4$ scale $\lambda$
(inverse of the size of the throat) in terms of the Planck
scale $L_P$:
$$ \lambda^{4} = {1 \over N L^4_P}, \eqno (20) $$
and inserts it into the geometric mean relations (15), one
obtains a value for the cosmological constant:
$$\Lambda_c = {1 \over {\sqrt N}} {1 \over L^4_P} = {M^4_P \over
{\sqrt N}}. \eqno (21) $$ which is a nice result because in the large $
N$
limit, this value of the cosmological constant is small.
The reason $AdS_4$ is relevant to estimate the cosmological constant (
vacuum energy density) is because it corresponds naturally to a $vacuum$
of
the orginal BF-CS-Higgs inspired-action:
$$ F^{ab}_{\mu\nu} = 0. ~~~ Torsion = F^{a5}_{\mu\nu} = 0. ~~~ \phi^5 =
v. ~~~\phi^a = 0. \eqno (22) $$
the solutions to (22) incoporate the $AdS_4$ spaces in a natural way.
Using the decomposition of the $SO(3,2)$ gauge fields (13) in the vacuum
equations (22) and eq-(10) one arrives at:
$$ F^{ab}_{\mu\nu} = 0 \Rightarrow F^{ab}_{\mu\nu} = R^{ab}_{\mu\nu} (
\omega (e)) + \lambda^2 e^a_\mu \wedge e^b_\nu = 0 $$
$$ d \omega + \omega \wedge \omega + \lambda^2 e\wedge e= 0
\Rightarrow R = - \lambda^2. \eqno (23) $$
which is a hallmark of $AdS_4$ space: spaces of constant negative scalar
curvature.
Hence the $ AdS_4$ space is a natural $vacuum$ of the theory associated
with the inspired BF-CS-Higgs model (4).
Based on this fact that $AdS$ spaces are natural vacuum
solutions of the MMCW action, we will discuss the Vasiliev
construction of a theory of massless higher spin fields
excitations of $AdS_4$ based on higher spin (higher rank
tensors) algebras whose spin ranges $ s = 2, 3,
4....\infty$; i.e higher spin massless fields propagating
in curved $AdS$ backgrounds [11]. This procedure does
$not$ work in Minkowski spacetime. There is an infinite
number of terms in this theory involving arbitrary powers
of $\lambda$. This bypasses the no-go theorems of writing
consistent interactions of higher spin fields (greater than
$ s = 2 $) in flat Minkowski spacetime.
Higher spin algebras have been instrumental lately [12]
in understanding deeper the Maldacena $AdS/$\-$CFT$ conjecture
and to construct $ N = 8 $ higher spin supergravity
theories in $ AdS_4$ which is conjectured to be the field
theory limit of $M$ theory on $ AdS_4 \times S^7 $. Based
on our BF-CS-Higgs action above one can find its
higher-spin extension using Vasiliev's procedure by
introducing a suitable noncommutative but associative star
product on an auxiliary ``fermionic'' phase space whose
deformation parameter (instead of the Planck constant
$\hbar$ in the conventional Moyal star product) is the
inverse length scale characterizing the size of $AdS_4$'s
throat $ \lambda = r^{-1} $. The ``classical'' $ \lambda =0$ limit is the
flat Minkowski space one.
The Vasiliev star product encoding the nonlinear and nonlocal higher spin
field dynamics is defined taking advantage of the local isomorphism of
the
algebras $ so(3,2) \sim sp(4, R)$ and has the same form as the Baker
integral product of the Moyal star product:
$$ (F*G) (Z, Y) = ({1\over 2 \pi} )^4\int d^2 u ~d^2 {\bar u}
~d^2 v ~ d^2 {\bar v} ~e^{i ( u^\alpha v_\alpha -
{\bar u}^{{\dot \alpha}} {\bar v}_{{\dot \alpha}})} F (Z+ Y, Y+U) ~G (Z -
V,
Y + V). \eqno (24) $$
where the spinorial coordinates are:
$$ Z^m = (z^\alpha, {\bar z}^{{\dot \alpha}}). ~~~Y_m = (y_\alpha,
{\bar y}_{{\dot \alpha}}). ~~~ \alpha, \beta = 1, 2.~~~ {\dot \alpha}
= 1, 2. \eqno (25) $$
The Vasiliev-algebra-valued one form $ W = dx^\mu W_\mu(x| Z, Y, Q)$
contains the master field that generates all the higher massless spin
gauge fields after a Taylor expansion:
$$W_\mu = \sum W_{\mu, \alpha_1 \alpha_2.....{\dot \beta_1} {\dot
\beta_2} ...} (x| Q)
z^{\alpha_1} z^{\alpha_2}....y^{{\dot \beta_1}}
y^{{\dot \beta_2}}.....\eqno (26) $$
where $ Q$ is a discrete set of Clifford variables (Klein
operators) that anticommute with the spinorial auxiliary
variables. The Vasiliev-algebra-valued field strengths are:
$ F (W) = dW + W * \wedge W$. The matter fields belong to
the Vasiliev-algebra-valued zero forms $\Phi (x| Z, Y, Q) $ and are the
generalization of the orginal Higgs scalars $ \phi^A$. There are
auxiliary
fields as well in order to have off-shell realizations of the Vasiliev
algebra and Stuckelberg compensating spinor-valued fields. The Vasiliev
generalization of our BF-CS-Higgs action (4) is:
$$ S^{*} = \int dY ~ dZ ~ \Phi* \wedge F(W) * \wedge F(W) + \Phi*
\wedge d \Phi * \wedge d \Phi * \wedge d \Phi * \wedge d \Phi. \eqno (27)
$$.
where the Higgs potential terms are:
$$ V^{*}_H = (\Phi * \Phi - v^2)* (\Phi * \Phi - v^2). \eqno (28)$$
One must add the terms in the action corresponding to the Stuckelberg and
auxiliary fields as well.
To our knowledge the auxiliary fields are still unkown at the present.
Without them, one cannot have an off-shell realization of the Vasiliev's
algebra that would allow us to construct the full action.
An integration of the above action w.r.t the auxiliary spinorial
coordinates will yield an effective
four-dimensional action in $ AdS_4$.
The main task will be to see whether or not such action furnishes the
well known higher spin equations of motion; i.e that action which encodes
the nonlinear higher spin dynamics of the infinite number of
higher spin massless gauge fields (including the spin two
graviton). This is beyond the scope of this work.
What one can verify directly in this work is based on the star product
deformations of the MMCW action ( 12 ):
$$ S^{*} = \int F(W) * \wedge F(W) \eqno ( 29 ) $$
where one performs an integration with respect to the variables $ x, Y, Z
$. A vacuum solution of the deformed action is:
$$ F(W) = 0 \Rightarrow dW + W * \wedge W = 0. \eqno ( 30 ) $$
which does agree with the well known higher spin equations of motion in $
d = 4 $ for the higher spin massless gauge fields.
One will have to include the equations of motion involving the matter and
the Stuckelberg compensating fields as well. To achieve this goal
requires using the more general action given by eqs-(27, 28), after
adding the terms involving the Stuckelberg and the auxiliary fields. A
covariant constancy condition, like the one appearing in eq-(9), is
consistent with the one in [ 11 ]. However, to show that the full
equations of motion ( involving all the fields ) follow from the deformed
action remains to be proven, mainly because the auxiliary fields are
unkown. The fact that the deformed action given in eq-( 29 ) yields the
vacuum equations ( 30 ), consistent with the massless gauge fields higher
spin equations of motion, is a nice starting point.
Recently, actions for Noncommutative Gravity, based on a different star
product, have been given by Chamseddine [ 1 5 ]. The star product
involves a higher derivative series expansion in terms of the quantity $
\theta^{\mu\nu} $ appearing in the commutator of two spacetime
coordinates:
$$ [ x^\mu, x^\nu ] = i \theta^{\mu\nu}. \eqno ( 31 ) $$.
The quantity $ \theta^{\mu\nu}$ is in general $ x $ dependent and one
must use the definition of the Kontsevich star product [ 16 ] which
differs from Vasiliev's star product. One can then perform a deformation
of the MMCW action ( 12 ) using such Kontsevich star product in the same
lines outlined by [ 15 ].  The main difference is that these
noncommutative actions in even dimensions require the use of $unitary$
groups and a complex metric, whereas the Vasiliev's star product is based
on symplectic groups and requires auxiliary spinorial coordinates.  In
both cases the metric and spin connection are suitable components of the
gauge fields. It is warranted to see is there is any connection between
both star product approaches for these deformed actions.  One is given in
powers of $\lambda $ and the other in powers of $\theta^{\mu\nu} $ ( in
general is proportional to the Planck scale ). For the discussion of the
plausible one-to-one correspondence between $W_{\infty}$ strings ( higher
conformal spins of an effective $ 3$-dim ``world-sheet''/membrane)
propagating on (the boundary of) $ AdS_4 \times S^7$ and Vasiliev higher
spin massless fields propagating on $ AdS_4 $ (whose boundary is $
S^1\times S^2 $) see [13]. It was pointed out how open $ W_{\infty}$
strings with $ SU(N)$ Chan-Paton factors ending on $ D$-branes may be
linked to Vafa and Bars $D = 12$-dimensional $F, S $ theory in the large
$N$ limit.

Mann et al [ 17 ] have investigated black hole solutions to $ 2 + 1 $
gravity coupled to topological matter with a vanishing cosmological
constant.  They found new features compared to ordinary Einstein gravity
( two possible new black holes ). It is important to see if  new black
hole solutions in $AdS$ spaces can be studied from our approach based on
BF-CS-Higgs models. The advantages of  using this approach, compared to
the MMCW actions ( 12 ),  is the role of the full symmetry $ so(3, 2 )
\sim sp ( 4, R )$ which is ammenable to a deformation via star products.
Its relation to Noncommutative Gravity [ 15 ] deserves further
investigation.

\bigskip
\centerline{ Acknowledgements}
\bigskip

We are indebted to J. Mahecha for his help in preparing the manuscript; to
F. Mansouri and M. Vasiliev for correspondence and to R. Mann for
reference [ 17 ].

\centerline {\bf References}

1- L. Freidel, K. Krasnov: ``$BF$ deformation of Higher dimensional
Gravity'' hep-th/9901069.
2- J. Baez: Lecture Notes in Physics, Springer Verlag {\bf 543} (2000)
25-94.
J. Baez: Class. Quant. Gravity {\bf 15} (1998) 1817.
3- A. Cattaneo, P. Cotta-Ramousino, F. Fulcito, M. Martellino, M.
Rinaldi,
A. Tanzini, M.Zeni: Comm. Math. Physics {\bf 197} (1998) 571.
4- N. Ikeda: ``Deformations of $BF$ theories and Topological Membranes''
hep-th/0105286.
5- M. Kontsevich: ``Deformation Quantization of Poisson Manifolds''
q-alg/9709040.
6- A. Cattaneo, G. Felder: math.QA/9902090.
7- N. Seiberg, E. Witten: JHEP 9909 (1999) 032; hep-th/9908142.
8- S. Ansoldi, C. Castro, E. Spallucci: Phys. Lett. {\bf B 504} (2001)
174.
S. Ansoldi, C. Castro, E. Spallucci: Class. Quan. Grav {\bf 18} (2001)
L-17-23.
S. Ansoldi, C. Castro, E. Spallucci: Class. Quant. Grav {\bf 18 } 2865.
C. Castro: ``Branes from Moyal Deformation quantization of Generlaized
Yang-Mills'' hep-th/9908115.
9- F. Wilczek: Physical Review Letters {\bf 80} (1998) 4851.
10-J. Maldacena: Adv. Theor. Math. Phys. {\bf 2} (1998) 231.
11- M. Vasiliev: ``Higher Spin Gauge Theories, Star Products and AdS
spaces'' hep-th/9910096.
12- E. Sezgin, P. Sundell: ``Higher Spin $ N = 8$ Supergravity in
$AdS_4$'' hep-th/9903020.
13- C. Castro: ``On the large $ N$ limit, $ W_{\infty}$ strings, Star
products.....'' hep-th/0106260
14- S. W. MacDowell, F. Mansouri: Phys. Rev. Lett { \bf 38} (1977) 739.
A. Chamseddine, P. West: Nuc. Phys. {\bf B 129 } (1977) 39.
F. Mansouri: Phys. Rev { \bf D 16 } (1977) 2456.
15- A. Chamseddine: " Invariant actions for Noncommutative Gravity "
hep-th/0202137.
16- M. Kontsevich: " Deformation quantization of Poisson Manifolds "
q-alg/9709040.
17 -J. Gegenberg, S. Carlip and R. Mann: Physical Review { \bf D 51 } (
1995 ) gr-qc/9410021.
\bye